\documentclass[12pt]{iopart}
\usepackage{iopams}
\usepackage{amsmath2}
\usepackage{graphicx}
\usepackage{color}
\usepackage{amssymb}

\begin{document}

\title[Theoretical concepts]
{Physics in one dimension: Theoretical concepts for quantum many body systems}

\author{K. Sch\"onhammer}

\address{Institut f\"ur Theoretische Physik, Universit\"at 
G\"ottingen, Friedrich-Hund-Platz 1, Germany}

\ead{schoenh@theorie.physik.uni-goettingen.de}

\begin{abstract}

Various sophisticated approximation methods exist for the
description of quantum many body systems. It was realized early
that the theoretical description can simplify considerably in
one dimensional systems and various exact solutions exist. 
 The focus in this introductory paper
is on fermionic systems and the emergence of the Luttinger liquid
concept.

\end{abstract}


\maketitle

\section{Introduction}
\label{sec:introduction}

The theoretical description of systems of interacting particles is
notoriously difficult and only few exact results are available.
Especially in the development of the theory of continuous phase
transitions it became obvious that the number $d$ of spatial dimensions
plays an important role.
While for most experimentally realizable systems one has $d=3$,
it early turned out that for $d=1$, i.e., one-dimensional systems
the theoretical description simplifies considerably and exact
results for various types of spin chains can be obtained. By 
a mapping to fermions this implies exact
results also for special lattice models of interacting ``electrons''
in one dimension as discussed in section 2. 

\noindent In this introductory paper the focus is on (normal) 
one-dimensional fermionic {\it quantum liquids},
which are many body systems in which the indistinguishability of the
elementary constituents is important. These particles can live on
a lattice or the continuous line.
 On low energy scales
the metallic state is a non-Fermi liquid characterized by a power law
decay of space-time correlation functions with interaction dependent
exponents. The name {\it Luttinger liquid} (LL) 
was termed for this
behaviour \cite {Haldane81}.

\noindent  While in the beginning of the theoretical developments
the corresponding models were considered a mere playground for
theoreticians recent developments have made the attempt
to experimentally verify Luttinger liquid behaviour a florishing
field of research as shown in this special issue.
 
\noindent  This paper gives a historical 
account of the emergence of the Luttinger liquid concept.
The typical properties were first found in the models   
proposed by Tomonaga \cite{Tomonaga} and Luttinger \cite{L},
which use rather restrictive assumptions about the interaction.
Tomonaga's important step was to realize and use the fact that
the low energy spectrum of noninteracting fermions in one dimension 
 is identical to
that of a harmonic chain. This allows to describe the interacting 
fermions as a system of coupled oscillators
\cite{Tomonaga}. Luttinger's calculation 
of the momentum distribution in the groundstate marks the appearance of power
laws for interacting fermions in one dimension \cite{L}.
 It was realized much later that
the low energy physics of these models is generic under rather weak
assumptions \cite{Haldane81,Giamarchi}.
 This low energy physics can be found also in 
bosonic many body systems  \cite{Giamarchi,BDZ}. Here the focus is on
the analytical description of fermionic systems. Important
computational techniques for one-dimensional quantum many body systems 
like the density matrix renormalization group (DMRG) \cite{White,Uli}
are not discussed here.

\section{Models}

In this section we present models which played an important role 
for the theoretical understanding of interacting quantum systems 
in one dimension. We begin with the anisotropic spin $1/2$ chain with nearest
neighbour interaction in an external field $h$. The Hamiltonian reads
\begin{equation}
\label{spinmodel}
H=\sum_i\left[ J_xs_i^xs_{i+1}^x+ J_ys_i^ys_{i+1}^y+J_zs_i^zs_{i+1}^z
-hs_i^z  \right ]~,
\end{equation}
with the exchange couplings $J_\alpha$.
The operators of the spin components on the same site $i$ obey the usual
angular momentum commutation relations and spin operators on
different sites commute. For $J_x=J_y=0$ one obtains the
{\it Ising chain} \cite{Ising}, the simplest model of
interacting spins. As all operators commute it can be considered a
classical spin model. Ising's exact solution
for the free energy showed that at finite
temperatures no symmetry breaking to a state with a finite 
magnetization occurs for vanishing external field.
It was suspected that this a special property of
 one dimension \cite{Peierls},
later  confirmed by Onsager's exact solution for
the free energy of the {\it two}-dimensional Ising model
with $h=0$, showing a phase transition at finite temperature \cite{Onsager}.
 
\noindent For the isotropic case  $J_x=J_y=J_z=J$ and putting $h=0$ 
one obtains the Hamiltonian of the Heisenberg spin
$1/2$ chain \cite{Heisenberg}. For $J<0$ the ground state is
ferromagnetically ordered, while for $J>0$ the spins are
antiferromagnetically correlated but not ordered in the groundstate.  
 In 1931 Bethe \cite{Bethe}
 presented his famous Ansatz for the exact eigenstates
and eigenvalues of this model which was later generalized
to a larger class of $1d$-models as discussed in various
 textbooks \cite{Gaudin,Giamarchi,EFGKK}.   
At the end of his paper Bethe announced a follow up paper whith a
generalization of his Ansatz to higher dimensions.
 The fact that it never appeared, again
shows that many body physics in one dimension is special. In the
following the sophisticated Bethe-Ansatz technique is not 
described, only exact results for Luttinger liquid parameters
for lattice models are mentioned later. Unfortunately no exact
results for the correlation functions discussed later are 
available within the Bethe Ansatz approach, but it should be mentioned
that significant progress has been made recently to calculate 
dynamical spin structure factors using a non-perturbative
(``form factor'') approach \cite{Jimbo,Korepin,Caux}. 

\noindent Another special case of the general spin model in
Eq. (\ref{spinmodel}) should be mentioned. For $J_y=J_z=0$ one
obtains the {\it transverse Ising model} \cite{Pfeuty} which
now serves as a standard model for a system with a quantum phase transition
\cite{Sachdev}. It was solved exactly  \cite{Pfeuty} by the use
of a  Jordan-Wigner transformation {\cite{JordanWigner,LSM} which relates
the set of spin $1/2$ operators to a set of spinless Fermi operators.
For $J_x=J_y$ the spin model in Eq. (\ref{spinmodel}) reads in the
 fermionic representation
\begin{equation}
\label{spinless}
H=\sum_i\left [\epsilon_0(c_i^\dagger c^{\phantom{\dagger}}_i-\frac{1}{2})
 -t (c_i^\dagger c^{\phantom{\dagger}}_{i+1}+H.c.) +U (c_i^\dagger
 c^{\phantom{\dagger}}_i-\frac{1}{2})  
 (c_{i+1}^\dagger c^{\phantom{\dagger}}_{i+1}-\frac{1}{2})    \right ]~,
\end{equation}
where $\epsilon_0=-h, t=-J_x/2$, $U=J_z$
and $c_i^{(\dagger)}$ is the annihilation (creation)
operator of a fermion at site $i$ . This Hamiltonian describes spinless
fermions on a chain with a nearest neighbour ``Coulomb interaction''.
The spin model with $J_z=0$ (``$XY$-model'') corresponds to
noninteracting fermions and can therefore be solved exactly \cite{LSM}.

\noindent The spinless model  in Eq. (\ref{spinless}) 
is one of the lattice models
which played an important role
in the emergence of the
Luttinger liquid concept. The other one is the one-dimensional Hubbard
model \cite{Hubbard,LiebWu,EFGKK}
\begin{equation}
\label{Hmodel}
H=\sum_{i,\sigma}\left [
 -t (c_{i,\sigma}^\dagger c^{\phantom{\dagger}}_{i+1,\sigma}+H.c.) +U
 c_{i,\uparrow}^\dagger c^{\phantom{\dagger}}_{i,\uparrow}  
  c_{i,\downarrow}^\dagger c^{\phantom{\dagger}}_{i,\downarrow}    \right ]~,
\end{equation}
where $\sigma$ is the spin label and $U$  the on-site Coulomb
repulsion.\\

In his seminal paper Tomonaga \cite{Tomonaga} treated interacting
fermions on the continuous line. 
Without impurities all fermionic models discussed in
this paper can be written in the form
\begin{equation}
\label{kspace}
H=\sum_k\epsilon_k c_k^\dagger c^{\phantom{\dagger}}_k+
\frac{1}{2}\sum_{k_1,k_2,k_3,k_4}   
v_{k_1 k_2 k_3 k_4}c_{k_1}^\dagger c_{k_2}^\dagger
c^{\phantom{\dagger}}_{k_4}
c^{\phantom{\dagger}}_{k_3}~,
\end{equation}
where $k$ is a double index $k,\sigma$ for models including spin like
the Hubbard model. For the lattice models 
Eq. (\ref{spinless}) and Eq. (\ref{Hmodel})
 the momenta $k_i$ are in  the first
Brillouin zone and for continuum models they are on the line extending
 from $-\infty$ to $\infty$. The energy dispersion $\epsilon_k$ and
the interaction matrix elements $ v_{k_1 k_2 k_3 k_4} $  are specified 
in the following sections.

\section{The Tomonaga-Luttinger model}

A decisive step towards an understanding of interacting
fermions in one dimension
beyond perturbation theory was Tomonaga's idea \cite{Tomonaga} 
 to {\it bosonize} the Hamiltonian
 Eq. (\ref{kspace}) for nonrelativistic particles on a line $L$
with periodic boundary conditions ($\hbar=1$)
\begin{equation}
\label{Tomo}
\epsilon_k=k^2/(2m)~,~~ v_{k_1 k_2 k_3
  k_4}=\frac{1}{L}\tilde v(k_1-k_3)\delta_{k_1+k_2,k_3+k_4}~.
 \end{equation}
Tomonaga  studied the case when the Fourier transform
of the two-body interaction $\tilde v(k)$ is nonzero only
for $|k|<k_c\equiv 2\pi n_c/L$, where the cut-off $k_c$ 
is much smaller than the the
Fermi momentum $k_F\equiv 2\pi n_F/L$.
This corresponds to a long range interaction in real space.
 Perturbation theory then indicates that the ground
state  and {\it low energy} excited states have negligible admixtures
of holes deep in the Fermi sea and particles with momenta $|k|-k_F\gg k_c$.
This is the motivation for Tomonaga's approximation to {\it linearize}
 the dispersion $\varepsilon_k$
 in the regions around the two Fermi points $\pm k_F$, with
particle-hole pairs present
\begin{equation}
k \approx \pm k_F : \quad \epsilon_k = \epsilon_F \pm v_F (k\mp
k_F),
\label{lin}
\end{equation}
with $v_F=k_F/m$ the Fermi velocity.
Tomonaga realized that the Fourier components of the operator of the
density 
\begin{equation}
\hat \rho_n =
 \int^{L/2}_{-L/2} \hat \rho(x)e^{-ik_nx} dx =\sum_{n'}
 c^\dagger_{n'}c^{\phantom{\dagger}}_{n'+n},
\label{Dichte}
\end{equation}
where $c^\dagger_{n'} (c^{\phantom{\dagger}}_{n'})$ creates (annihilates)
 a fermion in the state
with momentum $k_{n'} = 2\pi n'/L$, play a central role
not only for the interaction but also for
the {\it kinetic energy}. 
His important idea
was to split $\hat \rho_n$ for momentum transfer
 $|k_n|\ll k_F$ into two parts, one
containing operators of {\it ``right movers''} i.e. involving fermions
near the right Fermi point $k_F$ with velocity $v_F$ and {\it ``left
  movers''} involving fermions near $-k_F$ with velocity $-v_F$
\begin{equation}
\hat \rho_n = \sum_{n'> 0}c^\dagger_{n'}c^{\phantom{\dagger}}_{n'+n} 
+\sum_{n'\le
  0} c^\dagger_{n'} c^{\phantom{\dagger}}_{n'+n} \equiv \hat \rho_{n,+} 
+ \hat \rho_{n,-}~.
\label{Zerlegung}
\end{equation}
 In the subspace with {\it no holes
deep in the Fermi sea} in which all one-particle states $|k_j \rangle$ 
with  
 $|j|\le M=n_F-\gamma n_c$  are occupied, the
commutation relations \cite{Tomonaga}
\begin{equation}
[\hat\rho_{m,\alpha}, \hat\rho_{n,\beta}] = \alpha m \delta_{\alpha \beta}
\delta_{m,-n} \hat 1
\label{Komm3}
\end{equation}
hold for $|n|,|m|\le M$. The dimensionless constant $\gamma$ has 
to be chosen larger for stronger interaction.
If one defines the operators
\begin{equation}
b_n \equiv \frac{1}{\sqrt{|n|}}\left\{ \begin{array}{ll}
\hat\rho_{n,+} \hspace*{3ex} & \mbox{ for } n > 0\\
\hat\rho_{n,-} & \mbox{ for }n  < 0\end{array} \right.
\label{Boson1}
\end{equation}
and the corresponding adjoint operators $b^\dagger_n$   this leads  using
$\rho^\dagger_{n,\alpha} = \rho_{-n,\alpha}$ to the bosonic
commutation relations
\begin{equation}
[b_n, b_m] = 0, \quad [b^{\phantom{\dagger}}_n,b^\dagger_m] = 
\delta_{mn} \hat 1.
\label{Komm4}
\end{equation}
The kinetic energy of the right movers as well as that of the left
movers can be expressed as a bilinear
form of the $b^{(\dagger)}$-operators using a remarkable operator identity
first presented by Kronig in a different context
 \cite {Kronig,Voit95,Schoenh05}. For the right movers it reads
\begin{equation}
T_+=
\sum^\infty_{n=1}v_F k_n c^\dagger_n c_n=
v_F\frac{ 2\pi}{L}
 \left[ \sum^\infty_{m=1} m b^\dagger_m b^{\phantom{\dagger}}_m + \frac{1}{2}
{\cal N}_+ ({\cal N}_+ +1)\right],
\label{Kronig}
\end{equation}
where ${\cal N}_+=\sum^\infty_{n=1} c^\dagger_n
c_n^{\phantom{\dagger}}$
is the particle number of the right movers.
For its proof the commutation relations Eq. (\ref{Komm4}) have {\it
  not} to be used.

\noindent  As $\hat V$ is bilinear in
the $\hat\rho_n$ the same is true for the $\hat\rho_{n,\alpha}$.
For the linearized fermionic dispersion $\epsilon_k=v_F|k| +{\rm const.}$ shown
as the dashed line in Fig. 1 
and the two-body interaction in Eq. (\ref{Tomo})
the Hamiltonian in Eq. (\ref{kspace}) 
can  therefore apart from an additional term
{\it linear} in the particle number operators ${\cal N}_\pm$
 exactly be rewritten as
\begin{eqnarray}
\tilde H & = &
 \sum_{n>0} k_n \left\{
\left(v_F + \frac{\tilde v(k_n)}{2 \pi }\right)  
\left(b^\dagger_n b^{\phantom{\dagger}}_n
 + b^\dagger_{-n} b^{\phantom{\dagger}}_{-n}\right) \right. \nonumber
\\ && \left. +
\frac{ \tilde v(k_n)}{2 \pi}
\left(b^\dagger_n b^\dagger_{-n} + b_{-n} b_n\right)\right\}
+ \frac{ \pi}{2L}\left[v_N{\cal N}^2
 +v_J{\cal J}^2\right] \equiv  H_B +
H_{{\cal N,}{\cal J}},\quad\quad
\label{T1}
\end{eqnarray}
where ${\cal N}\equiv {\cal N}_+ + {\cal N}_-$
 is the total particle
number operator,
${\cal J}\equiv {\cal N}_+ - {\cal N}_-$ the ``current operator'',
 and the
velocities are given by $v_N = v_F + \tilde v(0)/\pi$ and $v_J =v_F$.

\noindent If one now {\it assumes} that the bosonic commutation
relations in Eq. (\ref{Komm4})
hold generally (see discussion below)
the operators $ H_B$ and $H_{{\cal N,}{\cal J}}$ commute and
with the Bogoliubov transformation
  $\alpha^\dagger_n =b^\dagger_n \cosh \theta_n -b_{-n} \sinh \theta_n$
to new boson operators the Hamiltonian $H_B$
can be brought into the form
\begin{equation}
H_B = \sum_{n \neq 0} \omega(k_n) \alpha^\dagger_n 
\alpha^{\phantom{\dagger}}_n + \;
\mbox{const}., ~~\omega(k_n) = v_F |k_n|\sqrt{1+\tilde v(k_n)/(\pi v_F)}
\label{T2} 
\end{equation}
and $\theta_n$ is determined by
\begin{equation}
 \tanh \theta_n=-\tilde v(k_n)/(2\pi v_F+\tilde v(k_n) ) .
\label{theta}
\end{equation}
For $|k_n|\ll k_c$ and a smooth $\tilde v(k)$
 the {\it boson dispersion} is approximately
linear $\omega(k_n)\approx v_c|k_n|$ with $v_c=\sqrt{v_Nv_J}$
the {\it charge velocity}. With the approximation to linearize $\epsilon_k$
around the Fermi points
it is strictly linear up to $k_c$
if $\tilde v(k)$ is constant up to the cut-off.

\noindent Besides the charge velocity $v_c$ the ``stiffness
constant'' $K\equiv \sqrt{v_J/v_N}$ plays an important role.
The noninteracting case yields $K=1$, attractive interactions $\tilde
v(0)<0$ lead to $K>1$ and repulsive interaction with $\tilde v(0)>0$
imply $0< K<1$.
 For the generalized
model, where $\tilde v(k_n)$ is replaced by $g_4(k_n) $ in the first
line on the rhs of Eq. (\ref{T1}) 
(scattering events on one of the Fermi points)
and by $g_2(k_n)$ in the second line (scattering events involving
 both Fermi points which conserve the number of right and left movers)
the two {\it independent} quantities $v_c$ and $K$ describe the low energy
physics. This generalization
to non-Galilei-invariant systems
 turns out to be important for the general
Luttinger liquid concept discussed in section 4.

\noindent A simple trick to extend the range of validity of
Eq. (\ref{Komm4})
and therefore for the step from Eq. (\ref{T1}) to Eq. (\ref{T2})
is to make the Fermi sea {\it deeper} for fixed $k_c$ by 
extending the left (right)
mover branch to positive (negative) $k$-values up to
 $k_{\rm Band}>0$ ($-k_{\rm Band}<0$)
as shown in Fig. 1 for the (arbitrary) value $k_{\rm Band}=1.5k_F$.
The Kronig relation is easily extended to this case and leads to an
additional term linear in the particle number operators and therefore
Eq. (\ref{T1}) still holds.
\begin{figure} [t]
\begin{center}
  \includegraphics[width=0.5\linewidth,clip]{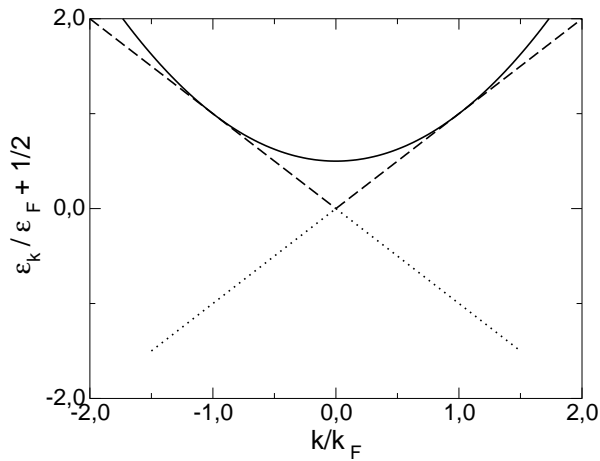}
\caption{Energy dispersion as a function of momentum. The full curve
shows the usual ``nonrelativistic'' dispersion and the dashed curve the
 linearized version.
The dot-dashed parts are the additional states for $k_{\rm Band}=1.5k_F$. The 
model discussed by Luttinger corresponds to $k_{\rm Band} \to \infty$.}
\end{center}
\label{Fig.1}
\end{figure}
Luttinger treated a model with two
{\it infinite} branches
of right and left moving fermions with dispersion $\pm v_F k$
\cite {L}. As he made an error related to the fact that his
Hamiltonian is not bounded from below, it is better to switch from
Tomonaga's to Luttinger's model keeping $k_{\rm Band}$ finite before
taking the limit  $k_{\rm Band}\to \infty$ \cite{Gutfreund}. Because of the
close relation of both models the term ``{\it Tomonaga-Luttinger (TL) model''}
 is often used.\\ 

 Fortunately Luttinger's error
had no influence on his inquiry if a discontinuity of  $\langle
n_{k,+}\rangle$ at $k_F$ exists
in the exact ground state of the interacting model,
as expected from Fermi liquid theory \cite{Landau,PN}.
After a 
lengthy calculation using properties of ``Toeplitz
determinants'' Luttinger found that the average occupation
  $\langle n_{k,+} \rangle$ in
the ground state for $k \approx k_F$  in the limit $L\to \infty$ 
behaves as
\begin{equation}
\label{nvonk}
\langle n_{k,+} \rangle -\frac{1}{2} \sim \left|\frac{k - k_F}{k_c}
\right|^{\alpha}
\mbox{sign} {(k_F-k)},
\end{equation}
where $\alpha\ge 0$ depends on the interaction strength (see below).
This is the power law behaviour mentioned in the introduction. It
cannot be obtained by finite order perturbation theory in the
interaction strength, which produces logarithmic terms in $|k-k_F|$.

\noindent Luttinger's error  was
corrected by Mattis and Lieb  \cite {ML} who presented a new algebraic method to
calculate   $\langle n_{k,+} \rangle$. They showed that
 Eq. (\ref{nvonk}) only holds for $\alpha<1$. For 
 $\alpha>1$ a term linear in $k - k_F$  dominates 
$\langle n_{k,+} \rangle -1/2 $. They also pointed out that one
obtains $ \langle n_{k,+} \rangle\equiv 1/2 $ in the limit
 interaction
cutoff $k_c\to \infty$ for a $k$-independent interaction.
In this limit Luttinger's model
 is equivalent to the massless Thirring
model \cite{Thirring} and can be 
written as a (quadratic) local bosonic field theory \cite{Haldane81,Giamarchi},
not discussed here further.\\

Additional information about the system is encoded in its time dependent
correlation functions $\langle A(t)B\rangle$, where   $\langle
\ldots \rangle$ denotes the expectation value in the ground state
(or in thermal equilibrium)
and $A(t)=e^{iHt}Ae^{-iHt}$ is the operator in the Heisenberg
picture. As Eq. (\ref{T2}) implies
$\alpha_n(t)=e^{-i\omega(k_n)t}\alpha_n$ the correlation function 
$\langle \hat \rho_n(t)\hat \rho_{-n}\rangle $ can easily be calculated using 
the inverse Bogoliubov transformation. Apart from a prefactor
its Fourier transform in time is the {\it dynamical structure factor} 
$S(q,\omega)$ \cite{AM} at $q=k_n$     . For the  Hamiltonian
Eq. (\ref{T2}) $S(q,\omega)$ is proportional to
$|q|\delta(\omega-\omega(q))$. 
For the nonrelativistic dispersion $\epsilon_k=k^2/(2m)$
this can only be asymptotically correct for $q\to 0$. 
This can already be seen from
the exact calculation of   $S(q,\omega)$ for  the noninteracting case.
As $\epsilon_{k_F+q}-\epsilon_{k_F}=qv_F+q^2/(2m)$ and 
$\epsilon_{k_F}- \epsilon_{k_F-q}=qv_F-q^2/(2m) $, for $0<q\ll k_F$
 the dynamical structure factor for fixed $q$ as function of frequency
 takes the form of a narrow box centered at
 $v_Fq$ of width $ q^2/m$ and height $\sim 1/q$.

\noindent In the calculation of one-particle Green functions the
Heisenberg operator $c_n(t)$ enters. It cannot be expressed using
the ``first step'' of bosonization introduced by Tomonaga \cite{Tomonaga}.   
Even time independent expectation values like the momentum distribution 
 $\langle n_{k,+}\rangle=\langle c_{k,+}^\dagger c_{k,+}\rangle$
 require an additional theoretical concept as
used by Luttinger \cite{L} and Mattis and Lieb \cite{ML}.

 The calculation of  $\langle n_{k,+} \rangle$ can be further
 simplified by {\it bosonizing the field operator}. This
 concept was introduced by Schotte and Schotte in the
 context of x-ray absorption from a core hole in the presence of
a Fermi sea
 \cite{Schotte2}. To the calculation of correlation functions of the
Tomonaga-Luttinger-model it was first applied by Luther and Peschel
\cite{Luther74}. Later subtleties of this second step of bosonization 
were addressed \cite{Haldane81,vonDelft,Schoenh05}. 

\noindent In this step the $c_{n\pm}$ are not bosonized
directly but the field operators
 $ \psi_\pm(x)$, e.g. for the right movers
\begin{equation}
 \psi_+(x)\equiv\frac{1}{\sqrt L}  \sum^\infty_{n = -\infty} e^{ik_nx}c_{n,+}~,
\label{aux}
\end{equation}
where the limit $k_{\rm Band}\to \infty$ apparently is performed first. The
commutation relations of  $ \psi_+(x)$ with the boson
operators $b_m$ and $b^\dagger_m$ imply the form
\begin{equation}
 \psi_+ (x) = \hat O_+(x) e^{i\phi_+^\dagger (x)}
e^{i\phi_+(x)},~~~~~i\phi_+(x) = \sum^\infty_{n = 1}\frac{e^{ik_nx}}{\sqrt{n}} b_n~,
\label{BFO1}
\end{equation}
where the {\it Klein operator} $\hat O_+(x) $
lowers the fermion number by one and 
 commutes with all boson
operators. Its explicit form has been discussed in 
detail \cite{Haldane81,vonDelft,Schoenh05}. It is not presented here,
 only the fact that  $\hat{O}_+ (x)$ and
$\hat{O}_-(x')$ anticommute is mentioned.
 
\noindent Using Eq. (\ref{BFO1}), the Bogoliubov transformation and
the Baker-Hausdorff formula,
$e^{A+B}=e^Ae^Be^{-\frac{1}{2}[A,B]}$ if the operators $A$
and $B$ commute with $ [A,B] $,
 it is straightforward to calculate ground state
 one-particle Green functions like $iG_+^<(x,t)\equiv
\langle \psi^\dagger_+(0,0) \psi_+(x,t)\rangle$ which enter the
description of photoemission. 
Using $\alpha_n|E_0\rangle=0$ one obtains \cite{Luther74}
\begin{equation}
 ie^{i\mu t}G^<_+(x,t) = 
\frac{e^{ik_F x} }{L} \exp\left \{ \sum^\infty_{n=1}  
\frac {1}{n} \left [e^{-i(k_n x-\omega_n t)}+
2\sinh^2 \theta_n\left (\cos{(k_n x)} e^{i\omega_n t}-1 \right) \right
]\right \}~,
\label{GK1}
\end{equation}
where $\mu$ is the chemical potential. Putting $t=0$ the momentum
distribution $\langle n_{k,+} \rangle$ is obtained by Fourier transformation.
In the limit $L\to \infty $  the $\theta_n$ can be expressed by
a continuous function $\theta(k_n)$ and
the equal time Green function 
$G^<(x,0)$  decays like $x^{-(1+\alpha)}$ for $x\gg 1/k_c$ with
$\alpha=2\sinh^2[\theta(0)]$, in contrast to $x^{-1}$ in the
noninteracting case. Therefore $\alpha$ which in Eq. (\ref{nvonk})
determines the behaviour of $\langle n_{k,+} \rangle$ near the Fermi
point $k_F$ is called the {\it anomalous dimension}. It can also be
expressed in terms of the stiffness constant $K$
\begin{equation}
 \alpha=2\sinh^2\theta (0)=(K-1)^2/2K .
\label{anomalous}
\end{equation}
For small interaction $\alpha$ is proportional to $\tilde v(0)^2$.
 Only for 
the special case $\tilde v(0)=0$ the anomalous dimension vanishes and
$\langle n_{k,+} \rangle$ has a discontinuity at $k_F$, the hallmark
of Fermi liquid theory \cite{Landau,PN}.
 In the generic interacting case one has $\alpha >0$
and Luttinger's power law  Eq. (\ref{nvonk}) holds for $\alpha <1$
(see Fig. 2).

 \begin{figure} [hbt]
\begin{center}
\includegraphics[width=0.4\linewidth,angle=-90]{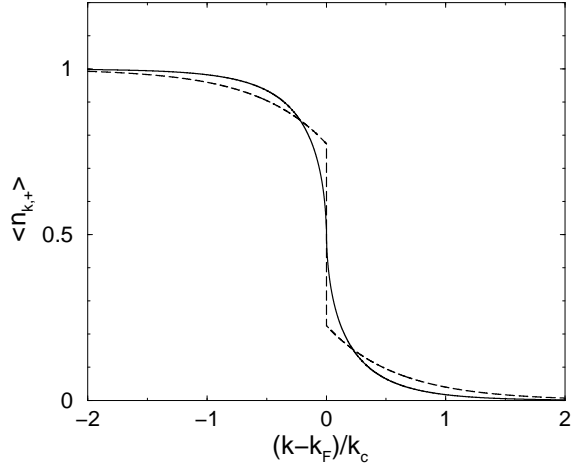}
\caption{ Average occupation $\langle n_{k,+} \rangle $ for 
different interactions specified by $\sinh\theta(k)$: 
The full line corresponds to $\sinh^2\theta(k)=0.3e^{-2|k|/k_c}$
i.e. the {\it finite} value $\alpha=0.6$. The expectation from Fermi liquid
theory with a finite jump at $k_F$ is shown as the dashed line for
 $\sinh^2\theta(k)=0.6(|k|/k_c)e^{-2|k|/k_c}$ i.e. {\it vanishing} anomalous
dimension. }
\end{center}
\label{Fig.Lutt}
\end{figure}
\noindent At finite temperatures $d\langle n_{k,+} \rangle/dk$
diverges like $T^{\alpha-1}$ at $k=k_F$ for $\alpha<1$. 

\noindent From the Fourier transform of $ie^{i\mu t}G^<(0,t)$ one obtains the
{\it local} spectral density $\rho^<(\omega)$, where $\omega$
is the energy relative to the chemical potential. 
For $T=0$ this leads to a power law suppression \cite{Luther74}
\begin{equation}
\rho^<(\omega)\sim \Theta (-\omega )\left (\frac{-\omega}{v_ck_c}\right )^\alpha
\label{local}
\end{equation}
of the spectral weight near the chemical potential
where $\Theta$ denotes the unit step function. This result holds
for arbitrary noninteger values of $\alpha\ge 0$
\cite{Meden}. The energy range over which this asymptotic behaviour
can be used depends on the functional form of $\tilde v(k)$.
 For finite temperatures the
local spectral weight is nonvanishing also for $\omega>0$ and
 $\rho^<(0)\sim T^\alpha$ in the low temperature limit \cite{Schoenh93b}.

\noindent The spectral function $\rho_+^<(k,\omega)$ relevant for describing
angular resolved photoemission is obtained from Eq. (\ref{GK1}) 
by a {\it double} Fourier transform.
Also the total spectral function  $\rho_+ (k,\omega)\equiv \rho_+^<(k,\omega) 
 +\rho_+^>(k,\omega)$ can be obtained from  $\rho_+^<(k,\omega)$
by using the relation $\rho_+^>(k_F+\tilde
k,\omega)=\rho_+^<(k_F-\tilde k,-\omega) $.

\noindent  For a general  $k$-dependence
of $\tilde v(k)$ the double transform can be performed analytically
only approximately \cite{Meden}. The exact calculation is
possible numerically, e.g. recursively \cite
{Schoenh93a,SM,Eggert}.
 An exception  
is provided by  $\rho_+^<(k_F,\omega)$. At the Fermi momentum 
one obtains for $\omega<0$
asymptotically 
 $\rho_+^<(k_F,\omega)\sim \alpha (-\omega)^{\alpha-1}$, i.e.
for $\tilde v(0)\ne 0$ a
power law  divergence as long as $\alpha <1 $. There is {\it no sharp
quasiparticle peak} as in a Fermi liquid. 
 This behaviour as well as the power laws in Eq. (\ref
{nvonk}) and Eq. (\ref {local}) are the hallmarks of Luttinger liquid
behaviour \cite{Haldane81}.

\noindent For $k_F-k_c<k<k_F$ and a {\it constant} $\tilde v(q)$ up to the
cutoff $k_c$ the $k$-resolved spectral function shows a power
law singularity at $\omega=v_c(k-k_F)$ if $\alpha<1/2$ \cite{Schoenh93a}
\begin{equation}
\rho^<_+(k_F+\tilde k,\omega) \sim \Theta (-\omega-v_c|\tilde k|)
(-\omega+v_c\tilde k)^{\frac{\alpha}{2} -1}(-\omega-v_c\tilde k )^{
\frac{\alpha}{2}}.
\label{RHOK2}
\end{equation} 
This result was first obtained assuming $2\sinh^2\theta(q)=\alpha
e^{-r|q|}$ and $\omega(q)=v_c|q|$ \cite{Luther74}. 
A comparison of Eqs. (\ref{T2}) and (\ref{theta})
shows that this is not consistent for all $q$.
 As the asymptotic analysis in two variables
 is less developed than for the one-dimensional
case it is not even known if for $\tilde k<0$
 a power law at $\omega=v_c\tilde k$
exists if the derivatives of $\tilde v(k)$ are different from zero
at $k=0$ \cite{Meden}. Another modification of
the power law singularities in Eq. (\ref{RHOK2})   for $k\ne k_F$
results from the corrections to the linearization of the
dispersion $\epsilon_k$ around the Fermi points 
\cite{Glazman1,Glazman2} as further discussed in section 5.\\

The TL model {\it including spin} was introduced and solved 
exactly using fermionic many-body techniques
by  Dzyaloshinski and Larkin \cite{DL73}. This included the
one-particle Green function $G_\pm(x,t)$. The implications 
for the spectral functions $\rho_\pm(k,\omega)$
when spin is included were not discussed.

\noindent Using bosonization the step from the spinless to the
spinful model is simple. All one has to do is to switch from the 
boson operators $b_{m,\sigma}^{(\dagger)}$ for the two spin components
$\sigma= \uparrow, \downarrow$ to ``charge'' $(c)$ and
 ``spin'' $(s)$ bosons \cite{Haldane81,Giamarchi}
\begin{equation}
b_{n,c}  \equiv \frac{1}{\sqrt{2}}~ (b_{n,\uparrow} + b_{n,\downarrow})~,~~~~
b_{n,s}  \equiv  \frac{1}{\sqrt{2}}~ (b_{n \uparrow} - b_{n,
  \downarrow})~~.
\label{Boson2}
\end{equation}
 One can write the TL-Hamiltonian $ H^{(1/2)}_{TL}$ 
for spin one-half fermions as  \cite{Haldane81,Giamarchi}
\begin{equation}
 H^{(1/2)}_{TL} = H_{TL,c} + H_{TL,s}~,
\label{TL2}
\end{equation}
where the $ H_{TL,a}$ are of the form of Eq. (\ref{T1})  but
 the interaction
matrix elements have the additional label $a$. The two terms on the
rhs of Eq. (\ref{TL2})  {\em commute}, i.e. the charge and spin
excitations are completely independent. This is usually called
{\it spin-charge separation} and is another hallmark of LL physics.
 The ``diagonalization'' of the two
separate parts proceeds exactly as before and the low energy
excitations are ``massless bosons'' $\omega_{n,a} \approx v_a |k_n|$ with
the {\em charge velocity} $v_c=(v_{J_c}v_{N_c})^{1/2}$ and the {\em spin
  velocity} $v_s= (v_{J_s}v_{N_s})^{1/2}  $. The corresponding
two stiffness constants are given by  $K_c=(v_{J_c}/v_{N_c})^{1/2}$
and $K_s= (v_{J_s}/v_{N_s})^{1/2}  $.
 The low temperature
thermodynamic properties 
of the TL-model
including spin
can be expressed in terms of the
 four velocities $v_{N_c}, v_{J_c}$, $v_{N_s}, v_{J_s} $
or the four quantities $v_c,K_c,v_s,K_s$. For spin rotation
invariant interactions $K_s=1$ holds \cite{Haldane81,Giamarchi}.

\noindent The one-particle Green functions of the spinful model
 are given by the
square root of the product of the charge and spin part which 
individually are of the form in Eq. (\ref{GK1}) \cite{DL73,Meden92,Voit93}.
 The anomalous dimension is given by
$\alpha= \sinh^2{\theta_c(0)}+\sinh^2{\theta_s(0)}\equiv
\alpha_c+\alpha_s$, where $\alpha_s$ vanishes in the spin rotation
invariant case.
Again the spectral function  $\rho_+(k,\omega)$ can be calculated
analytically in the low energy regime for a constant $\tilde v(k)$ up
to the cutoff $k_c$. It shows {\it two} power law
singularities \cite{Meden92,Voit93} for sufficiently small values of
the $\alpha_a$. For $\alpha_s=0$ the ``spin singularity'' is determined
by the exponent $(2\alpha-1)/2$ and the ``charge singularity'' by
$(\alpha-1)/2$.  These ``peaks'' disperse linearly with $k-k_F$.
For the modification of the singularities due to the $k$-dependence 
of $\tilde v(k)$ and the corrections to the linearisation of
$\epsilon_k$ the same arguments hold as in the spinless case.
At the Fermi momentum one again obtains   $\rho_+(k_F,\omega)\sim
|\omega|^{\alpha-1}$. For the local spectral density Eq. (\ref{local})
holds also in the spinful model, and the rhs of Eq. (\ref{nvonk})
also holds for $\langle n_{k\sigma,+}\rangle$.\\

Not all interesting results for 
correlation functions of the TL-model can be listed here.
 The unusual effect of
impurites on Luttinger liquids can e.g. be traced back to the 
$|Q|^{2(K-1)}$ divergence of the static density response function
at $k=\pm 2k_F+Q$ for repulsive interactions \cite{Luther74}.
This leads to the breakdown of a
perturbational analysis for an impurity potential
with a weak $\pm 2k_F$ backscattering contribution.
 The renormalization group analysis by
Kane and Fisher \cite{Kane92} showed that the backscattering potential
is a relevant perturbation for repulsive interactions
as expected from earlier work by Mattis \cite{Mattis}. The flow to
strong coupling implies that
the system behaves as if it is split by the impurity into two chains with
fixed boundaries at the end. Therefore it is necessary to
mention the behaviour of the one-particle Green function close to 
a boundary. The bosonization for periodic boundary conditions described
above has to be modified \cite{Fabrizio95}
to describe fixed boundary conditions. For spinless fermions
and $x$ close to the
boundary  $\langle \psi(x,0)\psi^\dagger (x,t)\rangle$ decays like
$(1/t)^{1+\alpha_B}$ in the long time limit, where $\alpha_B=1/K-1$
 is the boundary exponent. The local spectral function 
close to the boundary shows a power law $\rho (x,\omega)\sim
|\omega|^{\alpha_B}$,
where the proportionality factor contains an  oscillatory part in the
position variable.
The fact that in the low temperature limit
 the linear conductance of a backscattering impurity
vanishes like $T^{2\alpha_B}$ \cite{Kane92} can be understood
as an end-to-end tunneling between the ``split chains''.
    In contrast to the ``bulk'' anomalous dimension $\alpha$ the boundary 
value $\alpha_B$ is proportional to  $\tilde v(0)$ for small interactions.\\

A challenging problem is to describe the {\it disorder} in a Luttinger
liquid with a {\it finite impurity density} \cite{AR,GS,GMP}. We refer
to chapter 9 of Giamarchi's book for an extended discussion \cite{Giamarchi}.

\section{The Luttinger liquid concept}

Tomonaga was well aware of the limitations of his approach
for more generic
  two-body interactions 
(``In  the case of force of too short
range this method fails''\cite {Tomonaga}). It was only realized later
that the TL model is the fixed point Hamiltonian for a rather
general class of models \cite{So,Haldane81,Walter,Walter2}.
 This emergence of the general Luttinger
liquid concept is discussed in this section.

 \noindent In the opposite limit $k_c\gg k_F$
and  a $k$-independent interaction Tomanaga's continuum model
corresponds to a short range
interaction in real space. Then the low energy scattering processes
with momentum transfer $\pm 2k_F$ have to be included. They are usually
modeled by the {\it additional} ``$g_1$''-interaction term   
\begin{equation}
H^{(1)}_{\rm int}=\sum_{\sigma,\sigma'}
 \int
\left ( g_{1\Vert}\delta_{ \sigma, \sigma' }+ 
 g_{1\perp}\delta_{ \sigma,- \sigma' }\right )
\psi^\dagger_{+,\sigma}(x)  \psi^\dagger_{-,\sigma'}(x) 
    \psi_{+,\sigma'}(x)   \psi_{-,\sigma}(x)   dx.
\label{Hg1}
\end{equation}
Introducing a band cutoff S{\'o}lyom \cite{So} made a renormalization 
group (RG) study of this interaction at the one loop level.
 If the variable $s$ runs from zero to infinity
 in the process of integrating out 
degrees of freedom he obtained for spin-independent interactions
$ g_{i\Vert}= g_{i\perp}=g_i$ ($i=1,2$)
\begin{equation}
\frac{dg_1(s)}{ds}=  
-\frac{1}{\pi v_F}g^2_1(s)~,~~~~
\frac{dg_2(s)}{ds}= 
- \frac{1}{2\pi v_F}g^2_1(s) 
\label{RG1}
\end{equation}
with the solution $g_1(s)=g_1/[1+sg_1/(\pi v_F)] $, where $g_1$ is the
starting value. The $g_4$-interaction is not renormalized. 
 For $ g_1> 0$ the interactions flow to the {\it fixed line}
$g^*_1=0, g^*_2=g_2-g_1/2$ and the {\it fixed point Hamiltonian is a
TL-model} \cite{So}.
This  shows the generic importance of the 
TL-model for repulsive interactions. 

\noindent For $g_1<0$ the flow is to strong coupling.
 In order to 
 understand the strong coupling regime Luther and Emery 
 bosonized the additional interaction $H^{(1)}_{\rm int}$
in Eq. (\ref{Hg1}) and showed that  
 ``spin-charge separation'' also holds for this
 model  \cite {LE}. The charge part
stays trivial with massless charge bosons as the elementary
interactions.
They showed that for a particular value of $g_{1\Vert}$
 the exact solution for the spin part of
 the Hamiltonian is possible using refermionization.
The spectrum for the spin excitations is
{\it gapped}. It is generally believed that these properties 
of {\it Luther-Emery phases} are not restricted to the
solvable parameter values.\\

Strong coupling phenomena which lead to deviations from LL behaviour
can  occur in the {\it lattice models}
like the ones discussed in section 2,
 when for commensurate fillings {\it Umklapp processes} become important
\cite{Giamarchi}. Here important results for the models in Eqs.
(\ref{spinless}) and (\ref{Hmodel}) are presented.\\ 

 For the spinless fermions  Eq. (\ref{spinless})  
the parameters in Eq. (\ref{kspace})
 for a chain of $N$ sites with periodic boundary
conditions and $k$-values in the first Brilloin zone
are given by
\begin{equation}
\epsilon_k=-2t\cos{k}~,~~~~
v_{ k_1,k_2k_3,k_4}  
= \frac{2U\cos (k_1-k_3)}{N}
\sum_{m=0,\pm 1}\delta_{k_1+k_2,k_3+k_4+2\pi m} 
\label{Umklapp}
\end{equation}
The $m=0$ term on the rhs of Eq. (\ref{Umklapp}) represents
the direct scattering terms and the $m=\pm 1$
terms the Umklapp processes. 
It is a low energy process in the {\it half filled} band case
$k_F=\pi/2$ discussed here.
 Renormalization group analysis around the
noninteracting fixed point shows that the Umklapp terms
are strongly irrelevant 
which implies that the system is a LL for small $U>0$ \cite{Shankar}.
For $U\gg t>0$ charge density wave (CDW) order develops in which
every other site is occupied in order to avoid the Coulomb penalty.
The mapping to the spin model suggests that the transition
occurs at $U_c=2t$ as for $U_c>2t$ the Ising term dominates.
The exact Bethe ansatz solution confirms this and shows that
the model at half filling is a LL for $|U|<2t$. The Luttinger liquid
parameters can be obtained using a ground state property and 
the lowest charge excitation \cite{Haldane80}.
For repulsive interactions $U>0$ the value of
  $  K=\pi/[2 \arccos{ \left( - U/2t \right)}]$
 decreases monotonously 
from the noninteracting value
$K=1$ 
to $K=1/2$ for  $U=2t$, which corresponds to an anomalous dimension
$\alpha=1/4$. In order to reach smaller values than $1/2$ for $K$
the interaction has to have a longer
range in real space \cite{Giamarchi}. The limit $K\to 0$ is reached 
by the bare Coulomb interaction as $\tilde v(k)\sim \log{(1/|k|)}$ for
$k\to 0$ and the system is not a LL. The $4k_F$-harmonic of the
density-density correlation function shows a very slow decay almost
like in a Wigner crystal \cite{Schulz93}. 

\noindent The limit in which the lattice constant and the density 
go to zero corresponds
to the continuum limit. The interaction goes over to a contact
interaction. Because of the Pauli principle its effect vanishes and 
$K\to 1$. This limit is very different for the Hubbard model Eq. 
(\ref{Hmodel}) as the onsite interaction is between electrons with
{\it different} spins.\\

The energy dispersion $\epsilon_k$ for the Hubbard model
Eq. (\ref{Hmodel})
is the same as
in Eq. (\ref{Umklapp}) and the interaction matrix elements 
$v_{k_1\sigma_1,k_2\sigma_2,k_3\sigma_3,k_4\sigma_4}$ have the same
$m$-sum. The $k$-independent prefactor 
is proportional to
 $\delta_{\sigma_1\sigma_3}\delta_{\sigma_2\sigma_4}\delta_{\sigma_1,-\sigma_2}$ .
To show the difference to the spinless model the focus is again on
the half filled band case which is metallic for $U=0$. The limit $U\gg t$
is easy to understand.  Each site is singly occupied in order to avoid
the Coulomb penalty. In this limit the model can be mapped to the spin 
model Eq. (\ref{spinmodel}) with $J_x=J_y=J_z=4t^2/U$ and $h=0$, i.e.
the spin-$1/2$ Heisenberg antiferromagnet  which has 
gapless excitations  \cite{Gaudin,Giamarchi,EFGKK}.   
 In contrast there is a large gap $\Delta_c\approx U$
for excitations in the charge sector. As the model can be solved
exactly by a generalized Bethe ansatz approach
this {\it Mott-Hubbard} gap can be obtained exactly by solving 
Lieb and Wu's integral equations \cite{LiebWu,EFGKK}. It turns out to
be  finite
for {\it all} ~$U>0$.  It is exponentially
small $\Delta_c \approx (8t/\pi)\sqrt{U/t} \exp{(-2\pi t/U}$) for 
$0<U\ll t$.  This shows that the Umklapp term is no longer
irrelevant at the noninteracting fixed point. As the
Pauli principle does not influence electrons of opposite spin
the RG analysis shows that the Umklapp terms are {\it marginally relevant}
at the noninteracting fixed point \cite{Schulz,Giamarchi,EFGKK}.   

\noindent When the band is {\it not} half filled Umklapp
is {\it not} a low energy process and the Hubbard model is a Luttinger
liquid with $K_s=1$ for all $U>0$. The LL parameters $K_c$ and $v_a$ ($a=c,s$) 
can be obtained
by numerically solving Lieb and Wu's integral equations.   
 The results show that $K_c \to 1/2$ for
$n\to 0 $ as well as $n\to 1$ (half filling)
for {\it all} $U>0$ \cite {SchulzPRL,Schulz} .

\noindent  For results for
various correlation functions we refer to the textbook on the
one-dimensional Hubbard model \cite{EFGKK}.\\

The discussion of the two lattice models Eqs. (\ref{spinless}) and 
(\ref{Hmodel}) shows the general importance of the Luttinger liquid
concept for one-dimensional fermions with repulsive interaction.
 Only for half filling qualitative deviations occur.\\

As mentioned in the introduction the focus of this 
 paper is on
fermionic systems.  {\it Bosons} in one
dimension with repulsive interaction also behave as Luttinger liquids.
For them there is an essential difference between the noninteracting  
and the interacting system \cite{Giamarchi,CCGOR}.
 For the nonrelativistic dispersion
$\epsilon_k=k^2/(2m)$ the excitation spectrum
of the bosonic many body system is linear in $|k|$,
i.e. has the typical LL form only if a {\it finite} interaction is
present. In addition to the sound velocity the low energy physics
is described by the stiffness constant $K$ which again 
determines the large distance and long time behaviour of
correlation functions. In contrast to fermionic systems the 
noninteracting bosons correspond to the limit $K\to \infty$.

\noindent   Very versatile
systems to experimentally test the Luttinger liquid
behaviour for bosons in one dimension are ultracold gases
in strong optical lattices \cite{BDZ}.\\

Examples for the experimental realization of Luttinger liquid behaviour
in quasi-one-dimensional electronic sytems are presented in a
separate introductory paper \cite{expintro}. The comparison
of theory and experiment faces the problem
that strictly one-dimensional systems are a theoretical idealization.
Apart from this even the coupling to an experimental probe presents
a nontrivial disturbance of a Luttinger liquid.
The coupling between the chains in a very anisotropic $3d$ compound 
generally, at low enough temperatures, leads to true {\it long range
  order}.
The order develops in the phase for which the algebraic
decay of the corresponding correlation function
of the single chain LL is slowest \cite {Schulz}.
This can lead e.g. to charge density wave (CDW), spin density wave (SDW)
order or superconductivity. 
 Unfortunately the weak
coupling between several LLs or the coupling
of a  LL to a substrate is theoretically not very well
understood \cite {Giamarchi,Schoenh05}. The discussion could easily
fill a paper itself.\\

Carbon nanotubes are one-dimensional metallic systems when
the semimetallic carbon sheet is properly wrapped. In contrast to 
the systems discussed above {\it two} bands cross the Fermi level
like in a two-leg-ladder \cite{Giamarchi}. Within an approximate treatment
of the interaction terms it is possible to describe the system in
close analogy to the ``regular'' Luttinger liquids discussed so far
\cite{EggerGogolin,KBF}.  \\

One-dimensional metals having elementary excitations which propagate 
along the boundary of a two-dimensional system in one direction only 
are another type of Luttinger liquids. Wen introduced the concept
of {\it chiral} Luttinger liquids to describe the edge excitations
in the fractional quantum Hall states \cite{Wen,Chang}.

\section{Extensions of the Luttinger liquid concept and outlook}

The Luttinger liquid concept has recently been extended further in various
ways. An important step was to examine effects
 beyond Tomonaga's linearization  
Eq. (\ref{lin}) by including terms of order $(k-k_F)^2$ (or order
 $(k-k_F)^3$ for the half filled lattice
 models) \cite{Glazman1,Glazman2}.
This makes relaxation processes possible which do not exist in 
``linear'' Luttinger liquids.
 The interaction modifies e.g. the
``narrow box'' centered around $qv_F$ in $S(q,\omega)$ discussed in
sec. 3 and the $k$-resolved spectral functions $\rho(k,\omega)$ for $k\ne k_F$.
The  new ``nonlinear Luttinger liquid'' phenomenology makes contact to
methods developed for describing the x-ray edge singularity of core
hole spectra in the presence of a Fermi sea \cite{ND,OT}. The many-body
dynamics is described using effective models for 
mobile quantum impurities 
in a {\it linear} Luttinger liquid \cite{ZCG,Glazman1,Glazman2}.\\

The temperature dependence of the spectral functions was discussed
 only briefly so far. In the low temperature regime 
$k_BT\ll v_ck_c$  the power law behaviour
is smoothed out but the anomalous dimension can be recovered
using  the low energy scaling
relation   $ \rho^<(\omega,k_BT)=T^{\alpha}F(\omega/k_BT)$
\cite{Schoenh93b,Bock,Ralph}. A new scenario can result in the limit
that one of several intrinsic energy scales goes to zero. An example
is the half filled Hubbard model discussed in the previous section.
In the limit $U\to \infty$ the exchange coupling $J=4t^2/U$  goes to
zero and with it the spin velocity $v_s$. In the temperature 
range $J\ll k_BT\ll v_c/a_0$  called {\it spin-incoherent} Luttinger liquid
regime the one-particle spectral functions show qualitatively different
behaviour from that discussed in section 3 \cite{Matveev,Cheianov}.
Details can be found in a review by Fiete \cite{Fiete}.

\noindent {\it Helical} Luttinger liquids are 
realized in helical conductors
e.g. on the edges of topological insulators. 
Unlike the chiral Luttinger liquid a helical LL does not break
time reversal symmetry in these systems with strong spin-orbit interaction.
Helical LL
 exhibit spin-filtered transport where right movers carry spin up
and left movers spin down \cite{WBZ,XM}.

\noindent Yet another type of Luttinger liquid behaviour can be realized
in the presence of nuclear moments. A new ordered phase can result 
by the coupling of these moments to the conduction electrons
via the RKKY interaction \cite{BSL1}. 
The resulting low energy physics was dubbed {\it spiral} Luttinger
liquid by the authors.
A detailed comparison of the spectral properties of regular, helical
and spiral Luttinger liquids was presented recently \cite{BBS,Schuricht}.\\

 In recent years a very active field of theoretical research
is to generalize the description of
 one-dimensional quantum systems to conditions far from
thermal equilibrium. A typical example is a finite (interacting)
quantum wire which in the initial state is 
attached from the left and right to noninteracting leads
with differing chemical potentials $\mu_{L(R)}$ and temperatures
$T_{L(R)}$ \cite{JMS,MSHMS}. The theoretical description
 is usually done using the Keldysh
technique \cite{Keldysh1,Keldysh2}. For an even more general class of
nonequilibrium states where the initial states of the leads are not of
the grand canonical form a new bosonization technique has been
developed \cite{Mirlin} which uses concepts familiar from ``full
counting statistics'' \cite {LL} as well as the
 x-ray edge problem \cite{ND,OT}.\\

Obviously this short introduction cannot cover all the important 
contributions to the theory of one-dimensional quantum many body
systems. As we started the discussion with the spin 1/2 chain 
we end by mentioning that important insights into the physics
of isotropic antiferromagnetic chains of arbitrary spin were obtained using
methods of conformal field theory \cite{A1,A2,A3}. 

\section{Acknowledgements}

For useful comments on the manuscript the author would like to thank 
P. Dargel, R. Egger, T. Giamarchi, A. Honecker, C. Karrasch, V. Meden,
 W. Metzner,  A. Mirlin, and W. Zwerger.

{}

\end{document}